\documentclass{ws-ijmpe}

\begin{document}

\markboth{G. Y. Qin, A. Majumder, C. Gale}{Electromagnetic
Radiation from Broken Symmetries in Relativistic Nuclear
Collisions}

\catchline{}{}{}{}{}
%%%%%%%%%%%%%%%%%%%%%%%%%%%%%%%%%%%%%%%%%%%%%%%%%%%%%%%%%%%%%%%%%%%%

\title{ELECTROMAGNETIC RADIATION FROM BROKEN SYMMETRIES IN RELATIVISTIC NUCLEAR COLLISIONS }

\author{GUANG-YOU QIN,  CHARLES GALE}

\address{Department of Physics, McGill University, 3600 University Street\\
Montreal, QC, H3A 2T8, Canada}
%\\
% qing@physics.mcgill.ca}

\author{ABHIJIT MAJUMDER}

\address{Department of Physics, Duke University, Box 90305,
Durham, NC, 27708, USA}
%amajum@phy.duke.edu}

% \author{CHARLES GALE}
%
% \address{Department of Physics, McGill University, 3600 University Street\\
% Montreal, QC, H3A 2T8, Canada\\
% gale@hep.physics.mcgill.ca}

\maketitle

\begin{history}
\received{(received date)} \revised{(revised date)}
%\accepted{(Day Month Year)}
%\comby{(xxxxxxxxxx)}
\end{history}

\begin{abstract}
A new channel of direct photon production from a quark
gluon plasma (QGP) is explored in the framework of high-temperature QCD. This process appears at next-to-leading order, in the
presence of a charge asymmetry in the excited matter. The photon
production rate from this new mechanism is suppressed compared to the QCD
annihilation and Compton scattering at low baryon density but assumes importance in baryon-rich matter.
\end{abstract}

%%%%%%%%%%%%%%%%%%%%%%%%%%%%%%%%%%%%%%%%%%%%%%%%%%%%%%%%%%%%%%%%%%%
\section{Introduction}
%%%%%%%%%%%%%%%%%%%%%%%%%%%%%%%%%%%%%%%%%%%%%%%%%%%%%%%%%%%%%%%%%%%

The exploration of highly excited strongly interacting matter
produced in the collision of heavy nuclei~\cite{Harris:1996zx} has
now entered a phase of detailed study. Lattice QCD simulations of
such matter in equilibrium had predicted the
existence of a phase transition in the vicinity of  $T_c \sim 170$
MeV at vanishing chemical potential~\cite{Karsch:2003jg}. This was inferred from the
observed sudden rise of the scaled pressure and the entropy
density at this temperature.  The experimental results of the 
Relativistic Heavy-Ion Collider (RHIC) detector collaborations
have set a lower bound of about 5 GeV/fm$^3$ on the energy density
at a time $\tau=1$ fm/c in central Au+Au
collisions~\cite{RHIC_Whitepapers}. Based on Lattice QCD estimates,
these numbers place the temperature of such matter upwards of
$300$ MeV, well into the expected deconfined phase of QCD matter.
% At such temperatures, the system may no longer be described by a
% weakly interacting hadronic resonance gas and one expects the

In spite of  the temperatures reached, the produced matter demonstrates
considerable collective behaviour as witnessed by the large elliptic
flow observed. This has led to the assertion that the produced matter
is strongly interacting~\cite{Gyulassy:2004zy}. A description of
such collective behaviour within the picture of a weakly interacting
quasi-particle plasma of quarks and gluons, using methods based on
perturbative QCD, has so far not met with much success. This has spurned
attempts to describe the matter through phenomenological models
where the degrees of freedom are heavy quark and gluon quasi-particles and
a tower of  bound states of  quarks and gluons~\cite{Shuryak:2003ty}.
Flavor off-diagonal susceptibilities, measured  in lattice QCD simulations,
place severe constraints on the existence of such bounds states in the flavor sector~~\cite{Koch:2005vg}.
The exploration of the degrees of freedom in the unflavored sector
(\emph{i.e.}, pure glue degrees of freedom) can only be carried out by experimental probes.
In recent articles, a means to probe this structure through jet
correlations~\cite{Koch:2005sx} was considered. In these proceedings,
we report on the first attempt to probe the pure glue sector through its
possible electromagnetic signature.

Gluons do not carry electric charge, yet their interactions may
generate electromagnetic signatures if the medium is itself
electrically charged. The presence of a non-vanishing electric
charge or a net asymmetry between the quark and anti-quark
populations leads to an explicit breaking of charge conjugation
invariance by the medium. This allows for  new channels of photon
production at finite chemical potential ($\mu$) which were absent
at $\mu=0$ . The possibility of such rates was first pointed out
in Refs.~\cite{Majumder:2000jr,Majumder:2003vt}. In the current
effort, the spectrum of real photons from such processes will be
calculated and compared to the leading order rates of
Refs.~\cite{Kapusta:1991qp,Dumitru:1993vz,Vija:1994is}. All such
previous calculations of the photon rate at finite chemical
potential have tended to focus on processes which are
non-vanishing at $\mu_B=0$. The current effort extends the
computations of  the full photon production rates by the inclusion
of  new channels which arise solely at finite chemical potential.

%%%%%%%%%%%%%%%%%%%%%%%%%%%%%%%%%%%%%%%%%%%%%%%%%%%%%%%%%%%%%%%%%
\section{Formalism and Calculation}
%%%%%%%%%%%%%%%%%%%%%%%%%%%%%%%%%%%%%%%%%%%%%%%%%%%%%%%%%%%%%%%%%%

\begin{figure}[htb]
\begin{center}
\includegraphics[width=10cm]{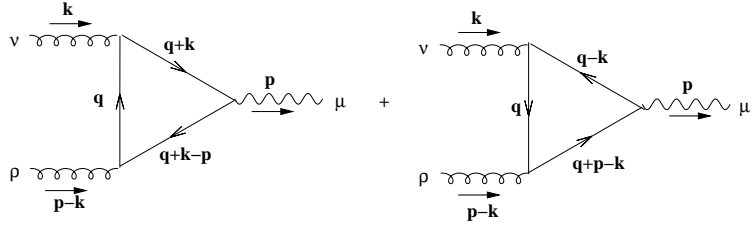}
\end{center}
\caption{The Feynman diagrams of gluon-gluon-photon vertex as the
sum of the two diagrams with quark numbers running in opposite
directions in the quark loops.} \label{gg-gamma}
\end{figure}

At zero temperature, as well as, at finite temperature and zero charge
density, diagrams in QED that contain a fermion loop with an odd
number of photon vertices (e.g. Fig. \ref{gg-gamma}) are cancelled
by an equal and opposite contribution coming from the same diagram
with fermion lines running in the opposite direction (Furry's
theorem~\cite{Furry:1939qr}). A physical perspective is obtained
by noting that all these diagrams are encountered in the
perturbative evaluation of Green's functions with an odd  number
of gauge field operators. At zero temperature, the focus lies on
quantities such as $\langle 0| A_{\mu_1} A_{\mu_2} ... A_{\mu_{2n+1}}
|0\rangle $, which vanishes under the action of the charge
conjugation operator $C$. At a temperature $T$, the corresponding
quantity to consider is
%%%%%%%%%%%%%%%%%%%%%%%%%%%%%%%%%%%%%%%%%%%%%%%%%%%%%%%%%%%%%%%%%
\begin{eqnarray}
Tr [ \rho(\mu,\beta)  A_{\mu_1} A_{\mu_2} ... A_{\mu_{2n+1}} ] =
\sum_{n} \langle n| A_{\mu_1} A_{\mu_2} ... A_{\mu_{2n+1}}
|n\rangle e^{-\beta (E_n - \mu Q_n)},
\end{eqnarray}
%%%%%%%%%%%%%%%%%%%%%%%%%%%%%%%%%%%%%%%%%%%%%%%%%%%%%%%%%%%%%%%%%
where $\beta = 1/T$ and $\mu$ is a chemical potential. In the case
of finite temperature and chemical potential, this quantity is
non-vanishing and Furry's theorem no longer holds. One may say
that the medium, being charged, manifestly breaks charge
conjugation invariance and these Green's functions are thus
finite, which leads to the appearance of new processes in a
perturbative expansion.

The above statement can be generalized almost unchanged to QCD for
processes with two gluons and an odd number of photon vertices.
The Feynman diagrams corresponding to the leading  contributions
to the new channel of photon production are those of Fig.~\ref{gg-gamma}~\cite{Majumder:2000jr}.
%%%%%%%%%%%%%%%%%%%%%%%%%%%%%%%%%%%%%%%%%%%%%%%%%%%%%%%%%%%%%%%%%%
%%%%%%%%%%%%%%%%%%%%%%%%%%%%%%%%%%%%%%%%%%%%%%%%%%%%%%%%%%%%%%%%%
At non-zero density, this leads to two new sources for photon
production: the fusion of gluons to form a photon ($gg\to\gamma$)
and the decay of a massive gluon into a photon and a softer gluon
($g\to g\gamma$). The full, physical, matrix element is obtained
by summing contributions from both diagrams,
$T^{\mu\nu\rho}(p,k,k') =
T_1^{\mu\nu\rho}(p,k,k')+T_2^{\mu\rho\nu}(p,k',k)$, where
$k'=p-k$. At finite temperature and chemical potential, the amplitudes may be expressed as:
%%%%%%%%%%%%%%%%%%%%%%%%%%%%%%%%%%%%%%%%%%%%%%%%%%%%%%%%%%%%%%%%%%
\begin{eqnarray}
T_1^{\mu\nu\rho}(p,k,k') &=& T\sum_{q_0} \int {d^3q\over (2\pi)^3}
eg^2 {\delta_{ab}\over 2} {\mathrm
Tr}[\gamma^\mu\gamma^\alpha\gamma^\nu\gamma^\beta\gamma^\rho\gamma^\gamma]
{(q+k)_\alpha q_\beta(q-k')_\gamma\over
(q+k)^2 q^2(q-k')^2}\nonumber \\
T_2^{\mu\rho\nu}(p,k',k) &=& T\sum_{q_0} \int {d^3q\over (2\pi)^3}
eg^2 {\delta_{ab}\over 2} {\mathrm
Tr}[\gamma^\mu\gamma^\gamma\gamma^\rho\gamma^\beta\gamma^\nu\gamma^\alpha]
{(q-k)_\alpha q_\beta(q+k')_\gamma\over (q-k)^2 q^2(q+k')^2}.
\end{eqnarray}
%%%%%%%%%%%%%%%%%%%%%%%%%%%%%%%%%%%%%%%%%%%%%%%%%%%%%%%%%%%%%%%%%
In the imaginary time formalism, the zeroth components of four
momentum are discrete Matsubara frequencies,
$q_0=i\omega_n+\mu=(2n+1)\pi T+\mu\ ,\ k_0=i\omega_k=2k\pi T\ ,\
p_0=i\omega_p=2p\pi T$, where integers $n$, $k$ and $p$ range from
$-\infty$ to $\infty$, and $\mu$ is the quark chemical potential.
The sum over the Matsubara frequencies may be conveniently
performed using the non-covariant propagator method of
Ref.~\cite{Pisarski:1987wc}. Here, one defines a
time-three-momentum propagator $\widetilde{\Delta}(\tau,E)$, as
%%%%%%%%%%%%%%%%%%%%%%%%%%%%%%%%%%%%%%%%%%%%%%%%%%%%%%%%%%%%%%%%%%
\begin{eqnarray}
\widetilde{\Delta} ({{{i\omega }}_n}\pm\mu ,E) &=& {{{{\int
}_0}}^{\beta }} d\tau {e^{{{{i\omega }}_n}\tau
}}{{\widetilde{\Delta} }_{\pm}}(\tau ,E).
\end{eqnarray}
%%%%%%%%%%%%%%%%%%%%%%%%%%%%%%%%%%%%%%%%%%%%%%%%%%%%%%%%%%%%%%%%%%
In the above equation, $E = |\vec{p}|$  represents the real energy
of the particle. The evaluation of the photon production
amplitudes as well as the ensuing rates are carried out in the
gauge invariant effective field theory of  QCD at high
temperature, the Hard-Thermal-Loops (HTL) effective
theory~\cite{Braaten:1989mz,KG}. In this formalism, one assumes the
condition that the temperature $T \rightarrow \infty$ and as a
result the coupling constant $g \rightarrow 0$. As a result a
hierarchy of scales emmerges \emph{i.e.}, $T >> gT >> g^2T$
\emph{etc.} Within this limit, one can derive effective
propagators and vertices for the soft modes $\sim gT$ by
integrating out the hard modes $\sim T$. In the current effort,
the effective two-gluon-photon vertex will be derived in this
limit.

\begin{figure}[htb]
\begin{center}
\includegraphics[width=4cm]{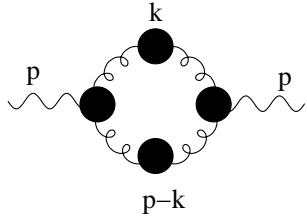}
\end{center}
\caption{The Feynman diagram of the photon self-energy evaluated
in this work, where the blobs represent either the effective
$gg\gamma$ vertices or the effective gluon
propagators.}\label{photon-SF}
\end{figure}

The production or absorption rate of photons from gluon fusion or decay in a dense medium
is related to the imaginary part of the photon self-energy depicted in Fig.~\ref{photon-SF}.
%%%%%%%%%%%%%%%%%%%%%%%%%%%%%%%%%%%%%%%%%%%%%%%%%%%%%%%%%%%%%%%%%%
%%%%%%%%%%%%%%%%%%%%%%%%%%%%%%%%%%%%%%%%%%%%%%%%%%%%%%%%%%%%%%%%%
In the case of a medium at equilibrium, the thermal photon
emission rate $R=d^4N/d^4x$ is related to the discontinuity or the
imaginary part of the retarded photon self-energy
$\Pi^R_{\mu\nu}$~\cite{KG},
%%%%%%%%%%%%%%%%%%%%%%%%%%%%%%%%%%%%%%%%%%%%%%%%%%%%%%%%%%%%%%%%%%
\begin{eqnarray}
E{dR\over d^3p} = {- g^{\mu\nu} \over (2\pi)^3} \mathrm{Im}
\Pi_{\mu\nu}^{R} {1\over e^{E\over T}-1}.
\end{eqnarray}
%%%%%%%%%%%%%%%%%%%%%%%%%%%%%%%%%%%%%%%%%%%%%%%%%%%%%%%%%%%%%%%%%%
The photon self-energy from the Fig.~\ref{photon-SF} may be
expressed formally as,
%%%%%%%%%%%%%%%%%%%%%%%%%%%%%%%%%%%%%%%%%%%%%%%%%%%%%%%%%%%%%%%%%%
\begin{eqnarray}
\Pi^{\mu\mu'}(p) = T\sum_{k_0}\int{d^3k\over
(2\pi)^3}T^{\mu\nu\rho}(p,k,k')S_{\nu\nu'}(k)
T'^{\mu'\nu'\rho'}(-p,-k,-k')S_{\rho\rho'}(k'),
\end{eqnarray}
%%%%%%%%%%%%%%%%%%%%%%%%%%%%%%%%%%%%%%%%%%%%%%%%%%%%%%%%%%%%%%%%%%
where $T^{\mu\nu\rho}(p,k,k')$ is the effective photon-gluon-gluon
vertex evaluated in the HTL limit  and $S_{\nu\nu'}(k)$ is the
effective gluon propagator, after summing up the HTL
contributions to the self-energy of the gluon. After summing over the
Matsubara frequency $k^0$ and evaluating the discontinuity across
the real $p^0$, the photon production rate may be cast into a
kinetic form,
%%%%%%%%%%%%%%%%%%%%%%%%%%%%%%%%%%%%%%%%%%%%%%%%%%%%%%%%%%%%%%%%%%
\begin{eqnarray}
E{dR\over d^3p} &=& \sum_{l=\pm}\sum_{i,j=\pm,0} {1\over
2(2\pi)^2} {1\over e^{E\over T}-1} \int{d^3k\over (2\pi)^3}
\nonumber\\&& \int d\omega \int d\omega'
\rho_i(\omega)\rho_j(\omega') \delta(\omega+\omega'-E)
(1+f(\omega)+f(\omega')) |M_{lij}|^2,
\end{eqnarray}
%%%%%%%%%%%%%%%%%%%%%%%%%%%%%%%%%%%%%%%%%%%%%%%%%%%%%%%%%%%%%%%%%%
where the matrix element
$M_{lij}=\epsilon_{l\mu}(p)\epsilon_{i\nu}(k)\epsilon_{j\rho}(k')T^{\mu\nu\rho}(p,k,k')$
and $\rho_i(\omega)$ and $\rho_j(\omega')$ are the spectral
functions of the longitudinal or transverse gluon propagators.
The product of two $\rho$ functions give three types of contribution:
pole-pole, pole-cut, and cut-cut. In this first effort, the focus
will lie on the hard photon production, \emph{i.e.}, photons with
momenta $p \sim T$, which requires that at least one of the gluons
in Fig.~\ref{photon-SF} to be hard. The cut-cut contribution with
two space-like gluons is dominant only in the region where both
gluon momenta are soft and is ignored in this effort.
%%%%%%%%%%%%%%%%%%%%%%%%%%%%%%%%%%%%%%%%%%%%%%%%%%%%%%%%%%%%%%%%%%
\section{Results}
\begin{figure}[htb]
%\vspace*{0.1cm}
\begin{center}
\includegraphics[width=8cm]{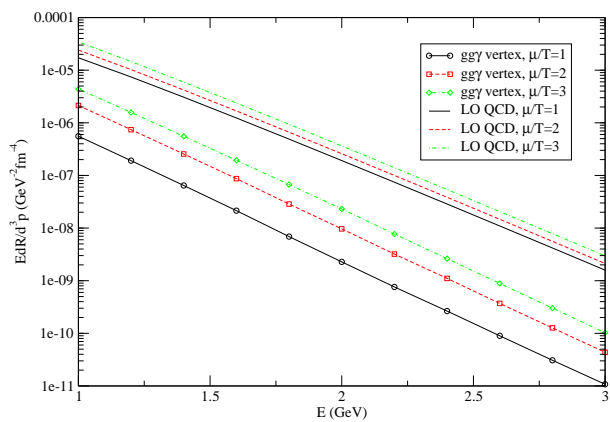}
\end{center}
\caption{The thermal photon emission rate from $gg\gamma$ vertex
in a medium at equilibrium with temperature $T=200\mathrm{MeV}$
compared with that from QCD annihilation and Compton processes.}
\label{T=200}
\end{figure}

%%%%%%%%%%%%%%%%%%%%%%%%%%%%%%%%%%%%%%%%%%%%%%%%%%%%%%%%%%%%%%%%%%
In this section, numerical results for the hard photon production
rate from a plasma with a finite charge density will be presented.
The calculation is performed for two massless flavors of  quarks
with $\mu_u=\mu_d=\mu=\mu_B/3$. In such a plasma, the strong
coupling constant is fixed, $\alpha_s=0.4$. The strange sector has
been ignored in this calculation.
In Fig.~\ref{T=200}, the photon
production from our new channel is compared with the contribution
from the leading order QCD processes of quark antiquark
annihilation and quark gluon Compton scattering.
%%%%%%%%%%%%%%%%%%%%%%%%%%%%%%%%%%%%%%%%%%%%%%%%%%%%%%%%%%%%%%%%%%
%%%%%%%%%%%%%%%%%%%%%%%%%%%%%%%%%%%%%%%%%%%%%%%%%%%%%%%%%%%%%%%%%
One may immediately note  that the
contribution from the  new channels, the $gg\gamma$ vertex, to the
photon production is much smaller than the QCD annihilation and
Compton processes at low chemical potential. However, with
increasing chemical potential at a fixed temperature, the photon
production rate from the $gg\gamma$ vertex tends to increase at a
swifter rate than QCD annihilation and Compton contribution. This
leads to the conclusion that in baryon-rich matter, such as that
produced in low energy collisions of heavy ions,
where the chemical potential of the medium is very
large, the new channel from $gg\gamma$ vertex will assume
significance in comparison to the leading order rates.

In the above estimates, the chemical potential ($\mu$) and temperature ($T$) are
held fixed separately. If the energy density were held constant, $T$ and $\mu$ would be related to each other by
the equation of state  (EOS). In this case, it has been shown that
the photon production rate from QCD annihilation and Compton
processes has a strong dependence on increasing chemical potential
$\mu$ of the medium~\cite{Dumitru:1993vz,Vija:1994is}.
Calculations of the photon production rate from the $gg\gamma$
vertex for the case of a constant energy density will be presented
in an upcoming publication~\cite{qin07}.

%%%%%%%%%%%%%%%%%%%%%%%%%%%%%%%%%%%%%%%%%%%%%%%%%%%%%%%%%%%%%%%%%%%
\section{Discussions and Conclusions}
%%%%%%%%%%%%%%%%%%%%%%%%%%%%%%%%%%%%%%%%%%%%%%%%%%%%%%%%%%%%%%%%%%%
In these proceedings, the hard photon signature emanating from a new
set of channels that arise solely when the medium explicitly
breaks charge conjugation invariance was presented. The photon production from
such channels is dependent on the gluon density of the medium and
thus offers a window to probe the gluon sector of the highly
excited strongly interacting matter. We employ an effective field theory 
of QCD  at high temperatures (HTL) and focus on the photon 
spectrum emanating from a medium at
equilibrium. In such a scenario, the photon rate from the new
channels is suppressed compared to the leading order rates for
realistic values of temperature and chemical potential, but will gain 
significance in baryon-rich matter. Our results may be easily
extended to non-equilibrium cases, such as the early stages of
relativistic nucleus-nucleus collisions, where the gluon
population far exceeds that of quarks; the photon production from
this new mechanism could outshine that from conventional channels.
Another interesting application of these rates is to the photon
production from jet plasma interactions~\cite{Fries:2002kt}.
Estimates of photon production from such diverse scenarios will
be explored in future efforts.

%%%%%%%%%%%%%%%%%%%%%%%%%%%%%%%%%%%%%%%%%%%%%%%%%%%%%%%%%%%%%%%%%%
\section{Acknowledgments}
%%%%%%%%%%%%%%%%%%%%%%%%%%%%%%%%%%%%%%%%%%%%%%%%%%%%%%%%%%%%%%%%%%

This work is supported in part by the U.S. Department of Energy
under grant DE-FG02- 05ER41367 and by the Natural Sciences and
Engineering Research Council of Canada.
%%%%%%%%%%%%%%%%%%%%%%%%%%%%%%%%%%%%%%%%%%%%%%%%%%%%%%%%%%%%%%%%%%

%%%%%%%%%%%%%%%%%%%%%%%%%%%%%%%%%%%%%%%%%%%%%%%%%%%%%%%%%%%%%%%%%%


\begin{thebibliography}{99}
%%%%%%%%%%%%%%%%%%%%%%%%%%%%%%%%%%%%%%%%%%%%%%%%%%%%%%%%%%%%%%%%%%


%\cite{Harris:1996zx}
\bibitem{Harris:1996zx}
  J.~W.~Harris and B.~M\"uller,
  %``The search for the quark-gluon plasma,''
  Ann.\ Rev.\ Nucl.\ Part.\ Sci.\  {\bf 46}, 71 (1996)
  %[arXiv:hep-ph/9602235].
  %%CITATION = HEP-PH 9602235


%\cite{Karsch:2003jg}
\bibitem{Karsch:2003jg}
  F.~Karsch and E.~Laermann,
  %``Thermodynamics and in-medium hadron properties from lattice QCD,''
  arXiv:hep-lat/0305025.
  %%CITATION = HEP-LAT 0305025;%%


%\cite{Wang:1996yf}
% \bibitem{Wang:1996yf}
%   X.~N.~Wang,
%   %``A pQCD-based approach to parton production and equilibration in
%   %high-energy nuclear collisions,''
%   Phys.\ Rept.\  {\bf 280}, 287 (1997)
  %[arXiv:hep-ph/9605214].
  %%CITATION = HEP-PH 9605214;%%


\bibitem{RHIC_Whitepapers}
  I.~Arsene {\em et al.},
  %Quark-Gluon Plasma and the Color Glass Condensate at RHIC?
  %The Perspective from the BRAHMS Experiment.
  Nucl.\ Phys.\ A {\bf 757}, 1 (2005);
  %nucl-ex/0410020
  B.~B.~Back {\em et al.},
  %The PHOBOS Perspective on Discoveries at RHIC.
  {\em ibid.} {\bf 757}, 28 (2005);
  %nucl-ex/0410022
  J.~Adams {\em et al.},
  %Experimental and Theoretical Challenges in the Search for the Quark
  %Gluon Plasma:  The STAR Collaboration's Critical Assessment of the
  %Evidence from RHIC Collisions.
  {\em ibid.} {\bf 757}, 102 (2005);
  K.~Adcox {\em et al.},
  %Formation of Dense Partonic Matter in Relativistic Nucleus-Nucleus
  %Collisions at RHIC:
  %Experimental Evaluation by the PHENIX Collaboration.
  {\em ibid.} {\bf 757}, 184 (2005).
  %nucl-ex/0410003


%\cite{Gyulassy:2004zy}
\bibitem{Gyulassy:2004zy}
  M.~Gyulassy and L.~McLerran,
  %``New forms of QCD matter discovered at RHIC,''
  Nucl.\ Phys.\ A {\bf 750}, 30 (2005).
%   [arXiv:nucl-th/0405013].
  %%CITATION = NUCL-TH 0405013;%%


%\cite{Shuryak:2003ty}
\bibitem{Shuryak:2003ty}
  E.~V.~Shuryak and I.~Zahed,
  %``Rethinking the properties of the quark gluon plasma at T approx. T(c),''
  Phys.\ Rev.\ C {\bf 70}, 021901 (2004);
%   [arXiv:hep-ph/0307267].
  %%CITATION = HEP-PH 0307267;%%
% %\cite{Shuryak:2004tx}
% \bibitem{Shuryak:2004tx}
%   E.~V.~Shuryak and I.~Zahed,
%   %``Towards a theory of binary bound states in the quark gluon plasma,''
  Phys.\ Rev.\ D {\bf 70}, 054507 (2004).
%   [arXiv:hep-ph/0403127].
  %%CITATION = HEP-PH 0403127;%%


%\cite{Koch:2005vg}
\bibitem{Koch:2005vg}
  V.~Koch, A.~Majumder and J.~Randrup,
%    ``Baryon-strangeness correlations: A diagnostic of strongly interacting
  %matter,''
  Phys.\ Rev.\ Lett.\  {\bf 95}, 182301 (2005).
%   [arXiv:nucl-th/0505052].
  %%CITATION = NUCL-TH 0505052;%%


%\cite{Koch:2005sx}
\bibitem{Koch:2005sx}
  V.~Koch, A.~Majumder and X.~N.~Wang,
  %``Cherenkov radiation from jets in heavy-ion collisions,''
  Phys.\ Rev.\ Lett.\  {\bf 96}, 172302 (2006);
%   [arXiv:nucl-th/0507063];
  %%CITATION = NUCL-TH 0507063;%%
%\cite{Majumder:2005sw}
% \bibitem{Majumder:2005sw}
  A.~Majumder and X.~N.~Wang,
  %``LPM interference and Cherenkov-like gluon bremsstrahlung in dense matter,''
  Phys.\ Rev.\ C {\bf 73}, 051901 (2006).
%   [arXiv:nucl-th/0507062].
  %%CITATION = NUCL-TH 0507062;%%


%\cite{Majumder:2000jr}
\bibitem{Majumder:2000jr}
  A.~Majumder and C.~Gale,
  %``Dileptons from a quark gluon plasma with finite baryon density,''
  Phys.\ Rev.\ D {\bf 63}, 114008 (2001); Phys. Rev. D {\bf 64}, 119901 (2001).
%   [arXiv:hep-ph/0011397].
  %%CITATION = HEP-PH 0011397;%%


%\cite{Majumder:2003vt}
\bibitem{Majumder:2003vt}
  A.~Majumder, A.~Bourque and C.~Gale,
%    ``Broken symmetries and dilepton production from gluon fusion in a quark
  %gluon plasma,''
  Phys.\ Rev.\ C {\bf 69}, 064901 (2004).
%   [arXiv:hep-ph/0311178].
  %%CITATION = HEP-PH 0311178;%%



%\cite{Kapusta:1991qp}
\bibitem{Kapusta:1991qp}
  J.~I.~Kapusta, P.~Lichard and D.~Seibert,
  %``High-Energy Photons From Quark - Gluon Plasma Versus Hot Hadronic Gas,''
  Phys.\ Rev.\ D {\bf 44}, 2774 (1991)
  [Erratum-ibid.\ D {\bf 47}, 4171 (1993)].
  %%CITATION = PHRVA,D44,2774;%%


%\cite{Dumitru:1993vz}
\bibitem{Dumitru:1993vz}
  A.~Dumitru, D.~H.~Rischke, T.~Schonfeld, L.~Winckelmann, H.~Stoecker and W.~Greiner,
  %``Suppression of dilepton production at finite baryon density,''
  Phys.\ Rev.\ Lett.\  {\bf 70}, 2860 (1993).
  %%CITATION = PRLTA,70,2860;%%

%\cite{Vija:1994is}
\bibitem{Vija:1994is}
  H.~Vija and M.~H.~Thoma,
  %``Braaten-Pisarski method at finite chemical potential,''
  Phys.\ Lett.\ B {\bf 342}, 212 (1995);
%   [arXiv:hep-ph/9409246];
  %%CITATION = HEP-PH 9409246;%%
%\cite{Traxler:1994hy}
%\bibitem{Traxler:1994hy}
  C.~T.~Traxler, H.~Vija and M.~H.~Thoma,
  %``Hard Photon Production Rate Of A Quark - Gluon Plasma At Finite Quark
  %Chemical Potential,''
  Phys.\ Lett.\ B {\bf 346}, 329 (1995).
%   [arXiv:hep-ph/9410309].
  %%CITATION = HEP-PH 9410309;%%



%\cite{Arnold:2001ba}
% \bibitem{Arnold:2001ba}
%   P.~Arnold, G.~D.~Moore and L.~G.~Yaffe,
%   %``Photon emission from ultrarelativistic plasmas,''
%   JHEP {\bf 0111}, 057 (2001);
% %   [arXiv:hep-ph/0109064];
%   %%CITATION = HEP-PH 0109064;%%
% %   P.~Arnold, G.~D.~Moore and L.~G.~Yaffe,
%   %``Photon emission from quark gluon plasma: Complete leading order  results,''
%   \emph{ibid.} {\bf 0112}, 009 (2001).
%   [arXiv:hep-ph/0111107].
  %%CITATION = HEP-PH 0111107;%%


%\cite{Furry:1939qr}
\bibitem{Furry:1939qr}
  W.~H.~Furry,
  %``On transition probabilities in double beta-disintegration,''
  Phys.\ Rev.\  {\bf 56}, 1184 (1939).
  %%CITATION = PHRVA,56,1184;%%

%\cite{Pisarski:1987wc}
\bibitem{Pisarski:1987wc}
  R.~D.~Pisarski,
  %``Computing Finite Temperature Loops With Ease,''
  Nucl.\ Phys.\  B {\bf 309} (1988) 476.
  %%CITATION = NUPHA,B309,476;%%

%\cite{Braaten:1989mz}
\bibitem{Braaten:1989mz}
  E.~Braaten and R.~D.~Pisarski,
  %``Soft Amplitudes in Hot Gauge Theories: A General Analysis,''
  Nucl.\ Phys.\  B {\bf 337}, 569 (1990);
  %%CITATION = NUPHA,B337,569;%%
%\cite{Frenkel:1989br}
% \bibitem{Frenkel:1989br}
  J.~Frenkel and J.~C.~Taylor,
  %``HIGH TEMPERATURE LIMIT OF THERMAL QCD,''
  Nucl.\ Phys.\  B {\bf 334}, 199 (1990).
  %%CITATION = NUPHA,B334,199;%%

\bibitem{KG}See, for example, Joseph I. Kapusta and Charles Gale, {\it Finite-Temperature Field Theory: Principles and Applications}, Cambridge University Press (Cambridge, 2006), and references therein.
%\cite{McLerran:1984ay}
%\bibitem{McLerran:1984ay}
%  L.~D.~McLerran and T.~Toimela,
  %``Photon And Dilepton Emission From The Quark - Gluon Plasma: Some General
  %Considerations,''
%  Phys.\ Rev.\ D {\bf 31}, 545 (1985).
  %%CITATION = PHRVA,D31,545;%%


%\cite{Gale:1990pn}
%\bibitem{Gale:1990pn}
%  C.~Gale and J.~I.~Kapusta,
  %``Vector dominance model at finite temperature,''
 % Nucl.\ Phys.\ B {\bf 357}, 65 (1991).
  %%CITATION = NUPHA,B357,65;%%

\bibitem{qin07} G.~Y.~Qin, A.~Majumder and C.~Gale, \emph{in preparation}.

%\cite{Fries:2002kt}
\bibitem{Fries:2002kt}
  R.~J.~Fries, B.~Muller and D.~K.~Srivastava,
  %``High energy photons from passage of jets through quark gluon plasma,''
  Phys.\ Rev.\ Lett.\  {\bf 90}, 132301 (2003).
%   [arXiv:nucl-th/0208001].
  %%CITATION = NUCL-TH 0208001;%%



%%%%%%%%%%%%%%%%%%%%%%%%%%%%%%%%%%%%%%%%%%%%%%%%%%%%%%%%%%%%%%%%%%
\end{thebibliography}
\end{document}